%% file: 14589.tex
\documentclass[letter,longauth,traditabstract]{aa}
\usepackage{graphicx}
\usepackage{multirow}
\usepackage{txfonts}
\begin{document}
   \title{{\it Herschel}-SPIRE FTS spectroscopy of the carbon-rich objects AFGL~2688, AFGL~618 and NGC~7027
\thanks{{\it Herschel} is an ESA space observatory with science instruments provided by European-led Principal Investigator consortia and with important participation from NASA.}}

   \author{
 R. Wesson \inst{1}\and
 J. Cernicharo \inst{2}\and
 M.J. Barlow\inst{1}\and
 M. Matsuura \inst{1,3}\and
 L. Decin \inst{4,5}\and
 M.A.T. Groenewegen \inst{6}\and
 E. T. Polehampton \inst{7,8}\and
 M. Agundez \inst{9} \and
 M. Cohen\inst{10}\and
 F. Daniel\inst{2}\and
 K. M. Exter\inst{4}\and
 W. K. Gear\inst{11}\and
 H. L. Gomez\inst{11}\and
 P. C. Hargrave\inst{11}\and
 P. Imhof\inst{12}\and
 R. J. Ivison\inst{13}\and
 S. J. Leeks\inst{7}\and
 T. L. Lim\inst{7}\and 
 G. Olofsson\inst{14}\and 
 G. Savini\inst{1}\and
 B. Sibthorpe\inst{13}\and
 B. M. Swinyard\inst{7}\and
 T. Ueta\inst{15}\and 
 D. K. Witherick\inst{1}\and
 J. A. Yates\inst{1}
}

   \institute{
Department of Physics and Astronomy, University College London, Gower Street, London WC1E 6BT, United Kingdom\and
Astrophysics Dept, CAB (INTA-CSIC), Crta Ajalvir km4, 28805 Torrejon de Ardoz, Madrid, Spain\and
Mullard Space Science Laboratory, University College London, Holmbury St. Mary, Dorking, Surrey RH5 6NT, UK\and
Instituut voor Sterrenkunde, Katholieke Universiteit Leuven, Celestijnenlaan 200 D, B-3001 Leuven, Belgium\and
Universiteit van Amsterdam, Sterrenkundig Instituut "Anton Pannekoek", Science Park 904, 1098 XH Amsterdam, Netherlands\and
Royal Observatory of Belgium, Ringlaan 3, B-1180 Brussels, Belgium\and
Space Science and Technology Department, Rutherford Appleton Laboratory, Oxfordshire, OX11 0QX, UK\and
Institute for Space Imaging Science, University of Lethbridge, 4401 University Drive, Lethbridge, Alberta, T1K 3M4, Canada\and
LUTH, Observatoire de Paris-Meudon, 5 Place Jules Janssen, 92190 Meudon, France\and
Radio Astronomy Laboratory, University of California at Berkeley, CA 94720, USA\and
School of Physics and Astronomy, Cardiff University, The Parade, Cardiff, Wales CF24 3AA, UK\and
Blue Sky Spectroscopy, 9/740 4 Ave S, Lethbridge, Alberta T1J 0N9, Canada\and
UK Astronomy Technology Centre, Royal Observatory Edinburgh, Blackford Hill, Edinburgh EH9 3HJ, UK\and
Dept of Astronomy, Stockholm University, AlbaNova University Center, Roslagstulsbacken 21, 10691 Stockholm, Sweden\and
Dept. of Physics and Astronomy, University of Denver, Mail Stop 6900, Denver, CO 80208, USA
}

   \date{}

 
  \abstract{We present far-infrared and submillimetre spectra of three carbon-rich evolved objects, AFGL 2688, AFGL 618 and NGC 7027. The spectra were obtained with the SPIRE Fourier transform spectrometer on board the {\it Herschel} Space Observatory, and cover wavelengths from 195-670$\mu$m, a region of the electromagnetic spectrum hitherto difficult to study in detail.  The far infrared spectra of these objects are rich and complex, and we measure over 150 lines in each object.  Lines due to 18 different species are detected.  We determine physical conditions from observations of the rotational lines of several molecules, and present initial large velocity gradient models for AFGL 618.  We detect water in AFGL~2688 for the first time, and confirm its presence in AFGL~618 in both ortho and para forms.  In addition, we report the detection of the J=1-0 line of CH$^+$ in NGC~7027.}
{}{}{}{}


   \maketitle
%

\section{Introduction}

Low- to intermediate-mass stars ($<$8M$_{\odot}$) shed much of their mass during the final stages of their evolution, in the form of a slow molecular wind.  The mass loss causes the star to leave the Asymptotic Giant Branch (AGB), moving to the left in the Hertzsprung-Russell diagram.  The star becomes hotter, warming and eventually ionising its previously ejected outer layers, which become visible as a planetary nebula (PN).  The ejecta from such stars are a major contributor to galactic chemical evolution (e.g. Matsuura et al. 2009).

The evolution from the AGB to the PN stage is rapid, and relatively few objects in this intermediate stage are known (326 are listed by Szczerba et al. (2007), compared to around 3000 known or suspected PNe (Frew \& Parker 2010)).  AFGL~618 and AFGL~2688 (the Egg Nebula) are two of the best-known objects in this transition stage.  AFGL~2688 is illuminated by a central star with spectral type F5, and Hubble Space Telescope images reveal a large number of round arcs crossed by `searchlight beams' (Sahai et al. 1998).  Proper motion measurements give a distance of 420pc and a dynamical age for the ejected material of 350 years (Ueta et al. 2006).  AFGL~618 entered the protoplanetary nebula (PPN) phase 100-200 years ago (Kwok \&  Bignell 1984, Bujarrabal et al. 1988), and is more evolved than AFGL~2688.  It contains a B0 central star surrounded by a compact H~{\sc ii} region (Wynn-Williams 1977, Kwok \&  Bignell 1984).  The variability of free-free emission from this object over the last 30 years implies rapid evolution (Kwok \&  Feldman 1981, S\'anchez Contreras et al. 2002).  NGC~7027 is a very young planetary nebula (Masson 1989).  It has been extensively studied due to its high surface brightness at all wavelengths, and observations with the Infrared Space Observatory (ISO) revealed a far-infrared spectrum rich in atomic and molecular lines (Liu et al. 1996).  Its ionised inner regions are surrounded by and partly obscured by a PDR and massive molecular envelope.

Previous far infrared and sub-mm observations of these evolving post-AGB objects, although limited in their spatial resolution and spectral coverage, have revealed rich spectra containing numerous molecular and dust features (e.g. Cox et al., 1996, Liu et al. 1996, Herpin et al. 2002, Pardo et al. 2004, 2005, 2007a, 2007b).  The {\it Herschel} Space Observatory (Pilbratt et al. 2010) now significantly extends our observational capabilities.  The SPIRE spectrometer covers wavelengths from 195-670$\mu$m, a region of the electromagnetic spectrum hitherto largely unexplored.  The SPIRE instrument, its in-orbit performance, and its scientific capabilities are described by Griffin et al. (2010), and the SPIRE astronomical calibration methods and accuracy are outlined by Swinyard et al. (2010).  Here we present observations of AFGL~2688, AFGL~618 and NGC~7027 using the SPIRE spectrometer.

\section{Observations}

Observations of AFGL~2688, AFGL~618 and NGC~7027 were obtained using the SPIRE Fourier transform spectrometer, which covers short (SSW, 194-313\,$\mu$m) and long (SLW, 303-671\,$\mu$m) wavelength bands simultaneously.  We combine data taken during the satellite's Performance Verification Phase with data taken during the Science Demonstration Phase for the Mass-loss of Evolved StarS (MESS\footnote{http://www.univie.ac.at/space/MESS/}) guaranteed time key program (Groenewegen et al., in preparation).  We reduced the spectra with the Herschel Interactive Processing Environment software (HIPE), using the standard point-source pipeline described by Swinyard et al. (2010).  A dark-sky observation taken close to the time of the object observations was subtracted to correct for telescope and sky emission, and the spectra were flux-calibrated using observations of the asteroid Vesta.  We combined three observations of AFGL~618, five of AFGL~2688 and seven of NGC~7027 to obtain our final spectra.  The total on-source exposure times were 11\,988s (AFGL~618), 4\,928s (AFGL~2688) and 11\,188s (NGC~7027).

The FTS produces spectra in which the line profiles are sinc functions; to eliminate the negative side-lobes this produces, we apodized the interferograms using the extended Norton-Beer function 1.5 (Naylor \& Tahic 2007).  This results in line profiles which are close to Gaussian, with a full width at half-maximum (FWHM) of 0.070cm\,$^{-1}$ (2.1\,GHz).  We then measured line intensities using the {\sc elf} Gaussian fitting routine in {\sc dipso} (Howarth et al. 2004).  We detected 203 lines in AFGL~2688, 214 in AFGL~618 and 164 in NGC~7027.  We present plots of the continuum-subtracted spectra in Fig.~1.

\begin{figure*}
  \centering
  \includegraphics[width=17.5cm]{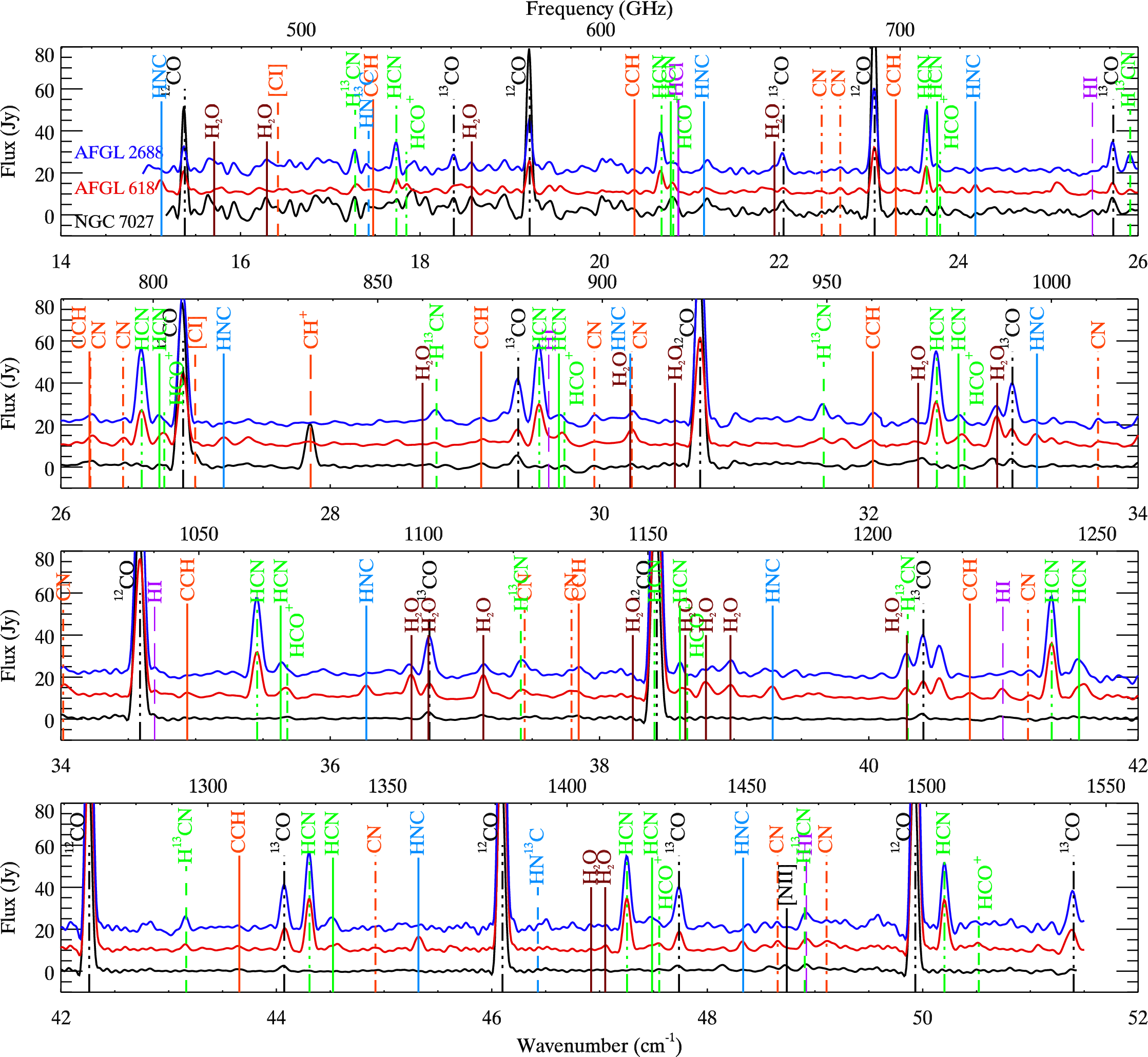}
\caption{Continuum-subtracted SPIRE FTS spectra of NGC~7027 (black), AFGL~618 (red) and AFGL~2688 (blue).  For clarity, the spectrum of NGC 7027 is multiplied by a factor of two, the spectrum of AFGL~618 is offset by 10\,Jy, and the spectrum of AFGL~2688 is offset by 20\,Jy.}
\end{figure*}

\section{Line measurements}

Most strong lines in our spectra are readily identified, although a number of lines remain unidentified at this stage.  Table~\ref{linelist}, available online, contains measured line intensities and identifications for each object spectrum; Tables~2, 3 and 4 list the strongest unidentified lines.  The quoted uncertainties are from the line fitting only; the absolute uncertainties on the flux calibration are estimated to be 10-20\% in the SSW band and $\sim$30\% in the SLW band.

\begin{figure*}

\centering
 
  \includegraphics[width=6cm]{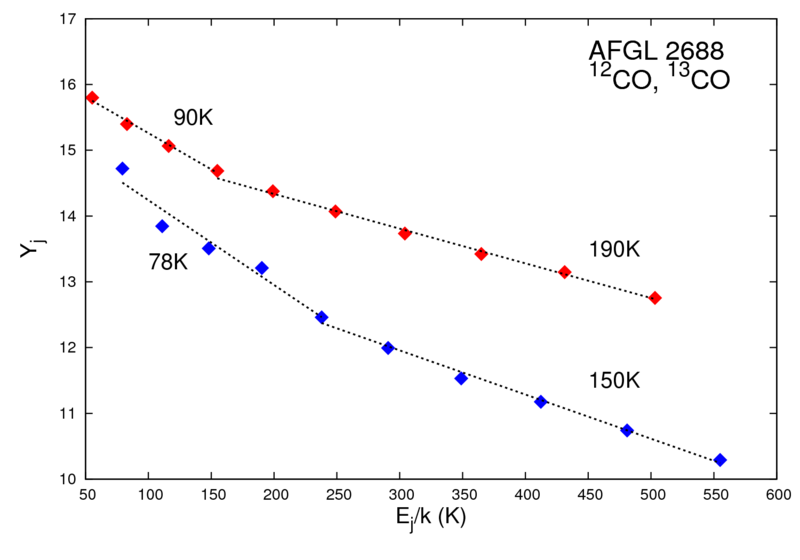}
  \includegraphics[width=6cm]{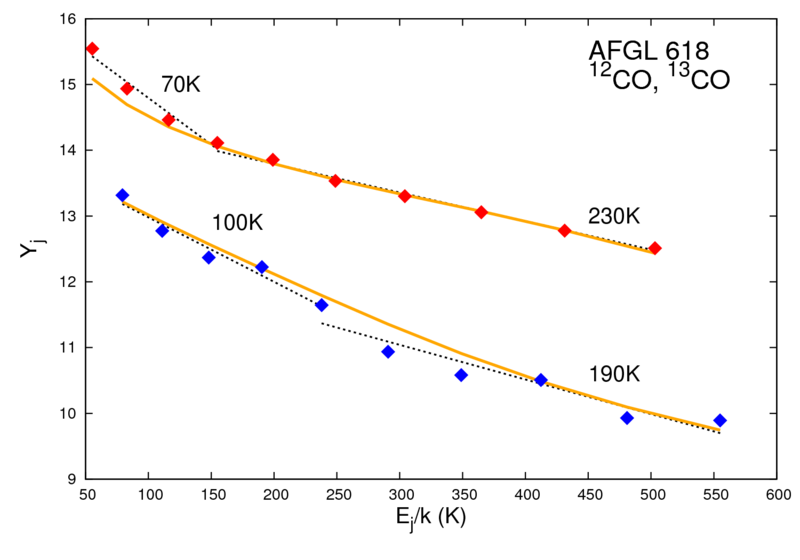}
  \includegraphics[width=6cm]{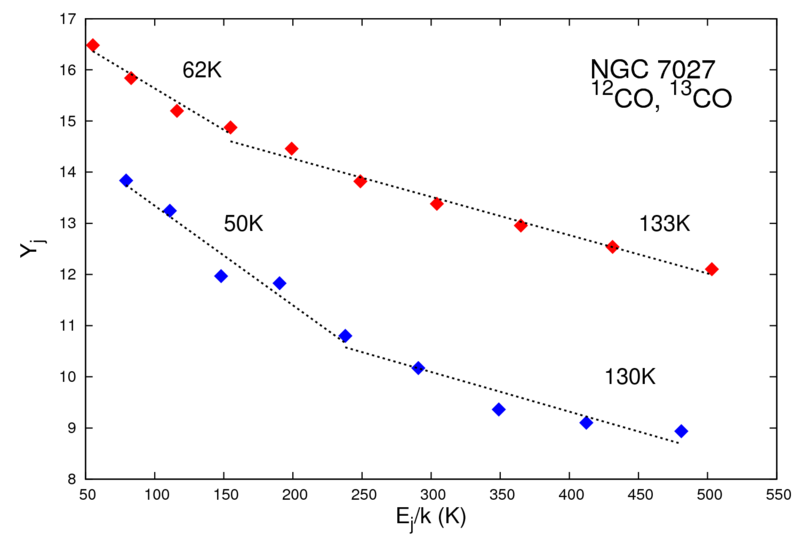}

  \includegraphics[width=6cm]{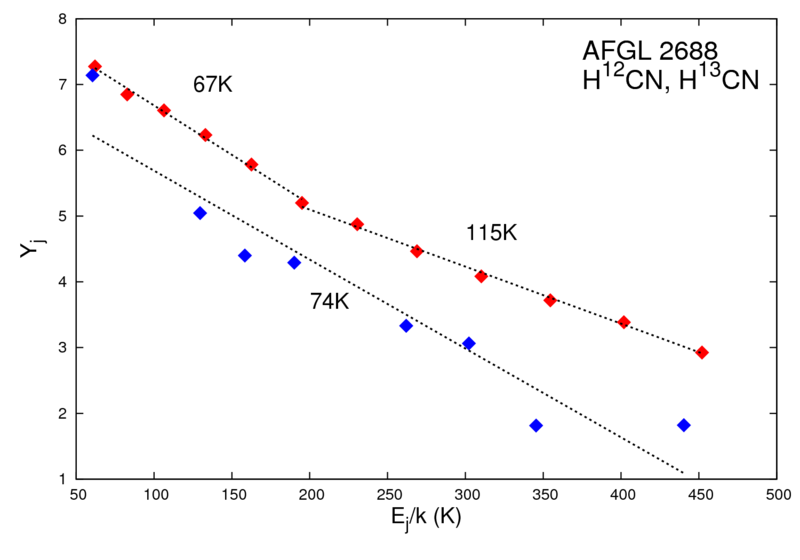}
  \includegraphics[width=6cm]{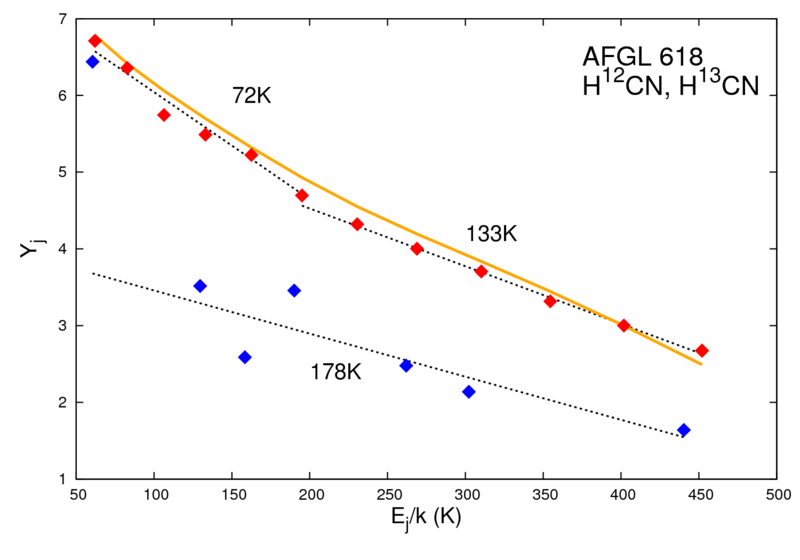} 
  \includegraphics[width=6cm]{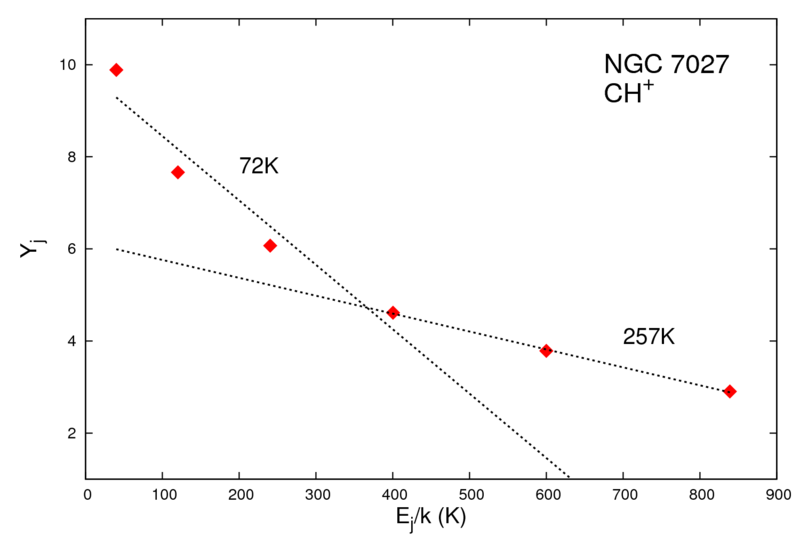}

\caption{Rotation diagrams based on fluxes measured in our SPIRE FTS spectra.  $Y_j = ln(4\pi F_{ji}/A_{ji}h\nu_{ji}g_{j})$, where F$_{ji}$ is the line flux in Wm$^{-2}$, A$_{ji}$ is the Einstein coefficient for spontaneous emission, $\nu_{ji}$ is the frequency of the transition and g$_j$ is the statistical weight of the upper level j.  The object and species are labelled in each panel.  Red points indicate species containing $^{12}$C, and blue points indicate $^{13}$C.  The temperatures derived from the slopes of the diagrams are indicated.  The CO, H$^{12}$CN and CH$^{+}$ data points clearly steepen at lower E$_{\rm j}$ and we fit these data with two components.  H$^{13}$CN data points are fitted with a single line.  The orange lines on the diagrams for AFGL~618 represent the predictions of our LVG models, described in Sect.~4.}
\label{excitation}
\end{figure*}

{\bf CO:}~All three spectra are dominated by the rotational lines of $^{12}$CO.  We plot rotational diagrams using the method described by Justtanont et al. (2000).  Figure~\ref{excitation} shows plots for $^{12}$CO and $^{13}$CO for our targets, using all the observed lines, from J=4-3 to J=13-12 for $^{12}$CO, and J=5-4 to J=14-13 for $^{13}$CO.  The CO J=10-9 line at 38.426\,cm$^{-1}$ is blended with HCN J=13-12 at 38.408\,cm$^{-1}$, and we estimate the relative contribution of the two species by linearly interpolating between the fluxes of the two neighbouring lines.  The sum of these estimated fluxes is within 3 per cent of the measured flux of the blend.

We derive rotational temperatures by fitting a line to the observed points, with the slope of the line being equal to 1/T$_{\rm {rot}}$.  The data clearly cannot be fitted by a single temperature, and so for $^{12}$CO we fit the data for J$_{\rm {up}\le}$7 and J$_{\rm {up}\ge}$8 separately, while for $^{13}$CO we fit the data for J$_{\rm {up}\le}$9 and J$_{\rm {up}\ge}$10 separately.  The temperatures we derive lie between 60 and 230K, and are shown on the plots in Fig.~\ref{excitation}.  The uncertainties on the rotational temperatures are typically 10-20\,K.

Justtanont et al. (2000) find considerably higher temperatures following this approach for lines up to J=37-36 observed in ISO-LWS spectra; plotting the new {\it Herschel} data together with the ISO data clearly shows that the CO lines cannot be described by a uniform rotational temperature, with the higher-J lines requiring higher temperatures.  This indicates that the low-J lines are tracing the extended AGB envelope, while the higher-J lines are tracing hotter gas from the inner layers of the outflow.  Models produced by Herpin \& Cernicharo (2000) and Herpin et al. (2002) use several spatial components at different temperatures to reproduce the observed emission from these objects.

{\bf HCN:}~We also plot rotation diagrams for H$^{12}$CN and H$^{13}$CN in AFGL~2688 and AFGL~618, where these species are well detected (Fig.~\ref{excitation}).  We derive rotation temperatures from fits to the observed points, using the interpolated flux described above for the J=13-12 transition of H$^{12}$CN, which is blended with the $^{12}$CO J=10-9 line.  The temperatures we derive are comparable to those derived from the $^{13}$CO lines, suggesting that these species trace similar regions of the circumstellar environments.

Rotational lines of vibrationally-excited HCN lines can be prominent in the AGB phase of carbon-rich stars (Cernicharo et al. 1996), and we detect them in AFGL~2688 and AFGL~618.  At the resolution of the SPIRE spectrometer, the l=-1 component of the $\nu_2$=1 level is blended with the ground state HCN lines, while the l=+1 component lines lie within one resolution element of those of HCO$^+$; ionised molecular species have very low abundances in AFGL~2688 and so we attribute lines at the frequencies of HCN $\nu_2$=1 to this species alone.  In AFGL~618, if the blend is not resolved we attribute the lines to both HCO$^{+}$ and HCN $\nu_2$=1.

{\bf H$_2$O:}~Herpin \& Cernicharo (2000) detected lines due to ortho-H$_2$O in AFGL 618 in ISO spectra.  We now also detect para-H$_2$O in this object, and we also detect water lines from AFGL~2688 for the first time.  We also detect seven low-lying water lines from NGC~7027.  The origin of oxygen-bearing molecules in carbon-rich objects is not well understood, given the expectation that all oxygen molecules will be locked up in CO with none left over to form oxygen-bearing molecules.  They may trace much older regions of the outflow from the star, from a time when its surface composition was oxygen-rich.  Cernicharo et al. (2004) showed that photodissociation of CO by the stellar UV field followed by reactions with H$_2$ can account for the observed abundance of H$_2$O in AFGL~618; AFGL~2688's much cooler central star makes a similar origin unlikely.  Alternative production mechanisms include the vaporization of icy bodies (Melnick et al. 2001), Fischer-Tropsch catalysis on grain surfaces (Willacy 2004) and radiative association between atomic oxygen and molecular hydrogen during the AGB phase (Ag\'undez \& Cernicharo 2006).  Another possibility that should be explored is how deep interstellar UV photons can penetrate into the envelope.  The circumstellar envelope of AFGL~2688 is formed by a series of concentric shells rather than a continuous envelope (Sahai et al. 1998), and if interstellar UV photons can reach its warm inner regions, then a very rich chemistry could be triggered by the photodissociation of CO and other species.  In the case of NGC~7027, water lines were tentatively identified in ISO-LWS spectra by Cernicharo et al. (1997) and Liu et al. (1997).  Yan et al. (1999) and Ag\'undez et al. (2010) showed that H$_2$O could form in NGC~7027's warm, dense PDR through reactions involving vibrationally-excited H$_2$.

{\bf CH$^{+}$:}~Lines due to ionised species are absent in the spectrum of AFGL~2688, weak in AFGL~618 and strong in NGC~7027.  The CH$^+$ line at 27.855\,cm$^{-1}$ is very bright in NGC~7027, and we combine our observation with LWS observations of higher-J CH$^{+}$ lines from Cernicharo et al. (1997) to produce the rotation diagram in Fig.~\ref{excitation}.  The data cannot be fitted by a single temperature, and we derive T$_{\rm {rot}}$=72$\pm$12\,K for E$_{\rm j}<$400\,K, and T$_{\rm {rot}}$=257$\pm$8\,K for E$_{\rm j}>$400\,K.  These temperatures are far lower than the $\sim$4300\,K activation energy of the reaction C$^+$~+~H$_2$~$\rightarrow$~CH$^+$~+~H.  However, under the simplifying assumptions of low optical depth and LTE, the derived rotational temperature is a lower limit to the kinetic temperature.   Ag\'undez et al (2010) find that CH$^{+}$ in NGC~7027 is formed both via this reaction in warm gas (T$_{\rm {\rm k}}>$400-500\,K) and via the reaction of vibrationally excited H$_2$ with C$^+$.

\section{LVG models for AFGL~618}

To further investigate the nature of AFGL~618, we constructed a large velocity gradient (LVG) model, based initially on that of Herpin \& Cernicharo (2000), in which the object is modelled as a torus surrounded by a high-velocity wind (HVW) region, with the ejecta from the earlier AGB phase further out.  Material in the torus is at temperatures from 250-1000\,K; the HVW region consists mostly of 200\,K material with a small amount of hotter gas at 1000\,K, and the AGB remnant has a temperature of 100\,K.  The CO rotation temperatures derived by Justtanont et al. (2000) from ISO data, and in this paper from {\it Herschel} data, are consistent with the model temperatures, with ISO sampling the warmer material and {\it Herschel} being sensitive to the cooler material further from the central star.  We used the model to predict $^{12}$CO, $^{13}$CO, H$^{12}$CN and H$^{13}$CN fluxes.

Matching our observed $^{12}$CO and $^{13}$CO fluxes was possible with almost exactly the same model parameters for CO as given in Table~1 of Herpin \& Cernicharo; the only change was to reduce the $^{13}$CO column density in the AGB ejecta from 3.5$\times10^{17}$\,cm$^{-2}$ to 2.5$\times10^{17}$\,cm$^{-2}$, reducing the total $^{13}$CO column density by 7\%.  This is well within the estimated model uncertainties of 20-30\%.  A good match to the observed H$^{12}$CN intensities required a total column density of HCN a factor of 5 lower than the value found by Herpin \& Cernicharo (2000).  This reflects the different regimes to which ISO and {\it Herschel} are sensitive, with {\it Herschel} providing information about much cooler regions than could be observed with ISO. The match to the H$^{13}$CN intensities is poor; these lines are quite weak in AFGL~618 and may be blended.  We plot the predictions of the model for $^{12}$CO, $^{13}$CO and H$^{12}$CN on the rotation diagrams in Fig.~\ref{excitation}.

The $^{12}$C/$^{13}$C ratio is a tracer of the history of nuclear processing inside stars.  $^{13}$C is an intermediate product in the CNO cycle which is the main fusion reaction converting hydrogen into helium in stars with masses$>$1.2M$_\odot$, and lower $^{12}$C/$^{13}$C ratios result from CNO cycle material being brought to the surface.  Milam et al. (2009) find that carbon-rich stars have $^{12}$CO/$^{13}$CO ratios generally in the range 25--90.  To match the observed CO line fluxes in AFGL~618 requires a $^{12}$C/$^{13}$C ratio of 21 in our model, implying substantial $^{13}$C enrichment.  Pardo et al. (2007a) find a $^{12}$C/$^{13}$C ratio of $\sim$15 in regions closest to the central star and $\sim$40 in regions further out, possibly indicating a late ejection of $^{13}$C-rich material from the star.

\section{Conclusions}

The unprecedented wavelength coverage and excellent sensitivity of the SPIRE spectrometer have allowed us to see the incredibly rich far-infrared and submm spectra of three carbon-rich evolved stars for the first time.  We have detected 150-200 emission lines in the SPIRE FTS spectra of AFGL~2688, AFGL~618 and NGC~7027, and presented preliminary investigations of the physical conditions in these objects based on the new data.  We find water in both ortho and para forms in all three objects, and observe strong ionised lines in the more evolved NGC 7027, including the J=1-0 line of CH$^{+}$.  This line, never previously detected in an astronomical source, is also detected in {\it Herschel} spectra of H~{\sc ii} region PDRs (Naylor et al. 2010).  LVG models of AFGL~618 are in excellent agreement with those based on earlier ISO data.  These observations show that the SPIRE spectrometer on board {\it Herschel} can provide significant new insights into the temperatures structure and chemical content of the outflows from evolved stars.

\begin{acknowledgements}

SPIRE has been developed by a consortium of institutes led by Cardiff Univ, (UK) and including Univ. Lethbridge (Canada); NAOC (China); CEA, LAM (France); IFSI, Univ. Padua (Italy); IAC (Spain); Stockholm Observatory (Sweden); Imperial College London, RAL, UCL-MSSL, UKATC, Univ. Sussex (UK); and Caltech, JPL, NHSC, Univ. Colorado (USA). This development has been supported by national funding agencies: CSA (Canada); NAOC (China); CEA, CNES, CNRS (France); ASI (Italy); MCINN (Spain); SNSB (Sweden); STFC (UK); and NASA (USA).  We thank Spanish MICINN for funding support through grants AYA2006-14876, AYA2009-07304 and the CONSOLIDER program "ASTROMOL" CSD2009-00038.  MATG and KME acknowledge support from the Belgian Federal Science Policy Office via the PRODEX Programme of ESA.  RW thanks Estelle Bayet for helpful discussions.

\end{acknowledgements}

\Online

\begin{table*}
{\tiny
\centering
\caption{Line intensities and identifications in each spectrum.  Rows beginning with a `+' indicate that the line is blended with that on the row above.  The quantum numbers given in the `transition' column are J except where stated.  H$_2$O transitions are given in the format J$^\prime$$_{k_a^\prime}$$_{k_c^{\prime}}$-J$_{k_a}$$_{k_c}$.}
\label{linelist}
\begin{tabular}{llll l l l l l l ll}
\hline
 & & & & \multicolumn{2}{c}{AFGL~2688} & \multicolumn{2}{c}{AFGL~618} & \multicolumn{2}{c}{NGC~7027}\\
$\bar{\nu}_0$ (cm$^{-1}$) & $\lambda_0$ ($\mu$m) & Species & Transition & $\bar{\nu}_{obs}$ (cm$^{-1}$) & F($\times$10$^{-18}$Wm$^{-2}$) & $\bar{\nu}_{ob s}$ (cm$^{-1}$) & F($\times$10$^{-18}$Wm$^{-2}$) & $\bar{\nu}_{obs}$ (cm$^{-1}$) & F($\times$10$^{-18}$Wm$^{-2}$) \\
\hline
\input{14589_t1a}
\hline
\end{tabular}
}
\end{table*}

\addtocounter{table}{-1}
\begin{table*}
{\tiny
\centering
\caption{-- continued.}
\label{linelist_full}
\begin{tabular}{llll l l l l l l ll}
\hline
 & & & & \multicolumn{2}{c}{AFGL~2688} & \multicolumn{2}{c}{AFGL~618} & \multicolumn{2}{c}{NGC~7027}\\
$\bar{\nu}_0$ (cm$^{-1}$) & $\lambda_0$ ($\mu$m) & Species & Transition & $\bar{\nu}_{obs}$ (cm$^{-1}$) & F($\times$10$^{-18}$Wm$^{-2}$) & $\bar{\nu}_{ob s}$ (cm$^{-1}$) & F($\times$10$^{-18}$Wm$^{-2}$) & $\bar{\nu}_{obs}$ (cm$^{-1}$) & F($\times$10$^{-18}$Wm$^{-2}$) \\
\hline
\input{14589_t1b}
\hline
\end{tabular}
}
\end{table*}

\begin{table}
\centering
\caption{Unidentified strong lines in the spectrum of AFGL~2688}
\begin{tabular}{llll}
\hline
$\bar{\nu}_0$ (cm$^{-1}$) & $\lambda_0$ ($\mu$m) & F($\times$10$^{-18}$Wm$^{-2}$) & Possible carriers \\
\hline
\input{14589_t2.tex}
\hline
\end{tabular}
\end{table}

\begin{table}
\centering
\caption{Unidentified lines in the spectrum of AFGL~618}
\begin{tabular}{llll}
\hline
$\bar{\nu}_0$ (cm$^{-1}$) & $\lambda_0$ ($\mu$m) & F($\times$10$^{-18}$Wm$^{-2}$) & Possible carriers \\
\hline
\input{14589_t3.tex}
\hline
\end{tabular}
\end{table}

\begin{table}
\centering
\caption{Unidentified lines in the spectrum of NGC~7027}
\begin{tabular}{llll}
\hline
$\bar{\nu}_0$ (cm$^{-1}$) & $\lambda_0$ ($\mu$m) & F($\times$10$^{-18}$Wm$^{-2}$) & Possible carriers \\
\hline
\input{14589_t4.tex}
\hline
\end{tabular}
\end{table}

\end{document}

%% file: 14589_t1a.tex
15.119 & 661.40 & HNC & 5-4 & & & 15.111 & 205.8$\pm$ 6.3&  & &  \\
15.239 & 656.23 & $^{12}$CO & $\nu$=1;4-3 & 15.248 & 63.0$\pm$ 8.1& & & \\ 
15.379 & 650.25 & $^{12}$CO & 4-3 & 15.372 & 281.4$\pm$ 8.7& 15.369 & 217.8$\pm$ 12.9&  15.371 & 556.5$\pm$ 6.0&  \\
15.707 & 636.65 & p-H$_2$O & 6$_{42}$-5$_{51}$ & & & 15.705 & 20.1$\pm$ 18.3&  & &  \\
16.294 & 613.71 & p-H$_2$O & 6$_{24}$-7$_{17}$ & & & 16.275 & 56.4$\pm$ 13.8&  16.282 & 129.9$\pm$ 21.9&  \\
16.417 & 609.14 & [C~{\sc i}] & 1-0 & & & & &  16.397 & 97.2$\pm$ 21.6&  \\
17.278 & 578.78 & H$^{13}$CN & 6-5 & 17.273 & 256.8$\pm$ 2.7& 17.293 & 127.5$\pm$ 9.0&  17.259 & 126.3$\pm$ 28.5&  \\
17.428 & 573.80 & HN$^{13}$C & 6-5 & 17.411 & 91.5$\pm$ 8.7& & &  17.420 & 60.0$\pm$ 13.8&  \\
17.478 & 572.16 & CCH & {\it N}=6-5 & & & 17.472 & 165.3$\pm$ 11.7&  & &  \\
17.736 & 563.82 & HCN & 6-5 & 17.733 & 325.8$\pm$ 6.6& 17.742 & 185.7$\pm$ 4.2&  17.748 & 31.8$\pm$ 11.4&  \\
17.848 & 560.30 & HCO$^+$ & 6-5 & & & 17.864 & 126.3$\pm$ 3.6&  & &  \\
18.377 & 544.16 & $^{13}$CO & 5-4 & 18.375 & 243.6$\pm$ 7.8& 18.364 & 59.7$\pm$ 9.0&  18.377 & 100.5$\pm$ 3.9&  \\
18.577 & 538.29 & o-H$_2$O & 1$_{10}$-1$_{01}$ & 18.571 & 164.4$\pm$ 3.9& 18.583 & 68.1$\pm$ 4.5&  18.574 & 154.5$\pm$ 14.7&  \\
19.222 & 520.23 & $^{12}$CO & 5-4 & 19.222 & 568.2$\pm$ 4.2& 19.223 & 358.2$\pm$ 1.5&  19.219 & 882.9$\pm$ 4.8&  \\
20.390 & 490.45 & CCH & {\it N}=7-6 & 20.384 & 39.6$\pm$ 2.4& 20.390 & 40.8$\pm$ 5.1&  20.374 & 42.3$\pm$ 4.8&  \\
20.691 & 483.30 & HCN & 7-6 & 20.687 & 456.9$\pm$ 3.3& 20.687 & 279.9$\pm$ 6.6&  20.674 & 43.5$\pm$ 2.7&  \\
20.793 & 480.93 & HCN & $\nu_2$=1;7-6f & 20.787 & 111.6$\pm$ 3.3& \multirow{2}{*}{20.818} & \multirow{2}{*}{150.6$\pm$ 6.9} &  & &  \\
20.821 & 480.28 & HCO$^+$ & 7-6 & & & & &  20.818 & 157.8$\pm$ 3.3&  \\
20.878 & 478.96 & HCl & 1-0 & 20.859 & 100.5$\pm$ 3.3& & &  & &  \\
21.165 & 472.48 & HNC & 7-6 & & & 21.176 & 89.4$\pm$ 4.5&  & &  \\
21.949 & 455.61 & p-H$_2$O & $\nu_2$=1;1$_{10}$-1$_{01}$ & 21.946 & 70.5$\pm$ 7.8& & &  & &  \\
22.051 & 453.50 & $^{13}$CO & 6-5 & 22.044 & 249.3$\pm$ 5.1& 22.054 & 85.5$\pm$ 3.0&  22.044 & 136.8$\pm$ 4.8&  \\
22.483 & 444.79 & CN & {\it N}=6-5;$\nu$=1 & 22.489 & 11.7$\pm$ 2.7& & &  & &  \\
22.683 & 440.86 & CN & {\it N}=6-5;$\nu$=0 & 22.681 & 73.5$\pm$ 5.7& & &  & &  \\
22.855 & 437.54 & $^{12}$CO & $\nu$=1;6-5 & 22.849 & 50.1$\pm$ 3.3& 22.872 & 45.6$\pm$ 38.1&  \\ 
23.065 & 433.56 & $^{12}$CO & 6-5 & 23.060 & 1002.0$\pm$ 3.6& 23.064 & 547.8$\pm$ 4.8&  23.058 & 1144.0$\pm$ 5.1&  \\
23.303 & 429.13 & CCH & {\it N}=8-7 & 23.309 & 58.5$\pm$ 3.9& 23.309 & 76.2$\pm$ 10.5&  23.306 & 33.9$\pm$ 3.6&  \\
23.646 & 422.91 & HCN & 8-7 & 23.643 & 693.9$\pm$ 4.5& 23.643 & 293.4$\pm$ 5.1&  23.636 & 40.8$\pm$ 3.6&  \\
23.762 & 420.84 & HCN & $\nu_2$=1;8-7f & 23.761 & 101.4$\pm$ 3.6& \multirow{2}{*}{23.789} & \multirow{2}{*}{118.2$\pm$ 6.0} &  & &  \\
23.795 & 420.26 & HCO$^+$ & 8-7 & & & & &  23.795 & 40.8$\pm$ 4.2&  \\
24.187 & 413.45 & HNC & 8-7 & 24.195 & 30.6$\pm$ 9.3& 24.191 & 110.7$\pm$ 1.8&  & &  \\
25.085 & 398.64 & o-H$_2$O & 2$_{11}$-2$_{02}$ & & & 25.113 & 228.3$\pm$4.5 \\
25.492 & 392.28 & H~{\sc i} & H20$\alpha$ & & & 25.507 & 50.1$\pm$ 3.9&  25.493 & 6.6$\pm$ 3.3&  \\
25.724 & 388.74 & $^{13}$CO & 7-6 & 25.720 & 380.1$\pm$ 7.5& 25.719 & 121.8$\pm$ 3.0&  25.720 & 81.6$\pm$ 4.5&  \\
25.912 & 385.92 & H$^{13}$CN & 9-8 & 25.914 & 235.2$\pm$ 6.9& 25.915 & 51.0$\pm$ 1.5&  & &  \\
26.171 & 382.10 & CN & {\it N}=7-6;$\nu$=1 & & & & &  26.167 & 28.5$\pm$ 4.2&  \\
26.226 & 381.30 & CCH & {\it N}=9-8 & 26.225 & 148.2$\pm$ 6.9& 26.233 & 147.0$\pm$ 3.0&  26.230 & 42.9$\pm$ 5.1&  \\
26.470 & 377.78 & CN & {\it N}=7-6;$\nu$=0 & 26.468 & 56.4$\pm$ 20.1& & &  & &  \\
26.600 & 375.95 & HCN & 9-8 & 26.596 & 855.0$\pm$ 9.6& 26.599 & 406.2$\pm$ 6.3&  & &  \\
26.731 & 374.10 & HCN & $\nu_2$=1;9-8f & 26.732 & 98.4$\pm$ 8.7& \multirow{2}{*}{26.756} & \multirow{2}{*}{166.2$\pm$ 6.6}&  & &  \\
26.767 & 373.59 & HCO$^+$ & 9-8 & & & & &  & &  \\
26.907 & 371.65 & $^{12}$CO & 7-6 & 26.905 & 1465.0$\pm$ 9.3& 26.907 & 825.9$\pm$ 6.3&  26.900& 1768.0$\pm$ 13.8&  \\
26.997 & 370.41 & [C~{\sc i}] & 2-1 & & & & &  26.997 & 113.4$\pm$ 11.1&  \\
27.208 & 367.54 & HNC & 9-8 & & & 27.211 & 113.7$\pm$ 2.1&  & &  \\
27.855 & 359.00 & CH$^+$ & 1-0 & & & 27.847 & 65.1$\pm$ 3.0&  27.851 & 466.5$\pm$ 4.5&  \\
28.685 & 348.61 & o-H$_2$O & $\nu_2$=1;2$_{11}$-2$_{02}$ & 28.677 & 60.9$\pm$ 2.1& 28.672 & 32.7$\pm$ 8.4&  & &  \\
28.789 & 347.36 & H$^{13}$CN & 10-9 & 28.781 & 207.3$\pm$ 2.7& 28.775 & 33.9$\pm$ 7.2&  & &  \\
29.121 & 343.39 & CCH & {\it N}=10-9 & 29.129 & 146.1$\pm$ 7.8& 29.139 & 91.8$\pm$ 8.7&  29.119 & 57.6$\pm$ 10.2&  \\
29.396 & 340.18 & $^{13}$CO & 8-7 & 29.391 & 543.9$\pm$ 5.4& 29.389 & 202.8$\pm$ 8.7&  29.386 & 136.8$\pm$ 12.6&  \\
29.553 & 338.38 & HCN & 10-9 & 29.550 & 916.8$\pm$ 5.7& 29.556 & 524.1$\pm$ 9.3&  & &  \\
29.622 & 337.59 & H~{\sc i} & H19$\alpha$ & & & & &  29.614 & 10.2$\pm$ 6.0&  \\
29.698 & 336.72 & HCN & $\nu_2$=1;10-9f & 29.695 & 172.8$\pm$ 6.9& \multirow{2}{*}{29.715} & \multirow{2}{*}{189.6$\pm$ 8.1}&  & &  \\
29.739 & 336.26 & HCO$^+$ & 10-9 & & & & &  29.735 & 7.8$\pm$ 5.7&  \\
29.970 & 333.67 & CN & {\it N}=8-7;$\nu$=1 & 29.974 & 117.3$\pm$ 13.2& & &  & &  \\
30.228 & 330.82 & p-H$_2$O & 9$_{28}$-8$_{35}$ & 30.244 & 167.7$\pm$ 13.5& & &  & &  \\
+30.229 & 330.81 & HNC & 10-9 & & & & &  & &  \\
+30.241 & 330.68 & CN & {\it N}=8-7;$\nu$=0 & \\
30.468 & 328.21 & $^{12}$CO & $\nu$=1;8-7 & & & 30.472 & 21.6$\pm$ 8.4&  & &  \\
30.560 & 327.22 & p-H$_2$O & 4$_{22}$-3$_{31}$ & & & 30.573 & 26.4$\pm$ 8.7&  & &  \\
30.748 & 325.23 & $^{12}$CO & 8-7 & 30.748 & 2073.0$\pm$ 16.5& 30.748 & 1229.0$\pm$ 10.2&  30.740 & 2249.0$\pm$ 11.7&  \\
31.665 & 315.80 & H$^{13}$CN & 11-10 & 31.653 & 297.3$\pm$ 24.0& 31.651 & 129.3$\pm$ 19.5&  & &  \\
32.030 & 312.21 & CCH & {\it N}=11-10 & & & 32.023 & 69.9$\pm$ 8.1&  & &  \\
32.366 & 308.96 & p-H$_2$O & 5$_{24}$-4$_{31}$ & & & 32.368 & 81.0$\pm$ 3.6&  & &  \\
32.505 & 307.64 & HCN & 11-10 & 32.501 & 818.7$\pm$ 6.3& 32.501 & 554.1$\pm$ 3.6&  & &  \\
32.665 & 306.13 & HCN & $\nu_2$=1;11-10f & 32.665 & 150.3$\pm$ 7.8& \multirow{2}{*}{32.703} & \multirow{2}{*}{129.3$\pm$ 10.5} &  & &  \\
32.711 & 305.71 & HCO$^+$ & 11-10 & & & & &  32.702 & 38.7$\pm$ 2.7&  \\
32.954 & 303.46 & p-H$_2$O & 2$_{02}$-1$_{11}$ & 32.946 & 236.4$\pm$ 7.2& 32.952 & 340.2$\pm$ 3.6&  32.943 & 44.1$\pm$ 3.3&  \\
33.067 & 302.41 & $^{13}$CO & 9-8 & 33.066 & 457.2$\pm$ 7.5& 33.066 & 202.5$\pm$ 3.9&  33.060 & 87.0$\pm$ 3.3&  \\
33.249 & 300.76 & HNC & 11-10 & & & 33.244 & 142.2$\pm$ 3.6&  & &  \\

%% file: 14589_t1b.tex
33.711 & 296.64 & CN & {\it N}=9-8;$\nu$=1 & & & 33.708 & 33.6$\pm$ 32.4&  & &  \\
34.026 & 293.90 & CN & {\it N}=9-8;$\nu$=0 & 34.030 & 124.8$\pm$ 13.8& 34.039 & 230.7$\pm$ 14.4&  & &  \\
34.588 & 289.12 & $^{12}$CO & 9-8 & 34.585 & 2711.0$\pm$ 17.4& 34.586 & 1586.0$\pm$ 10.2&  34.579 & 2114.0$\pm$ 12.0&  \\
34.695 & 288.23 & H~{\sc i} & H18$\alpha$ & & & & &  34.691 & 28.2$\pm$ 8.7&  \\
34.940 & 286.20 & CCH & {\it N}=12-11 & & & 34.945 & 75.9$\pm$ 6.0&  & &  \\
35.457 & 282.03 & HCN & 12-11 & 35.456 & 907.8$\pm$ 8.7& 35.457 & 521.7$\pm$ 4.2&  & &  \\
35.632 & 280.65 & HCN & $\nu_2$=1;12-11f & 35.639 & 237.6$\pm$ 14.4& \multirow{2}{*}{35.663} & \multirow{2}{*}{153.6$\pm$ 4.8}&  & &  \\
35.681 & 280.26 & HCO$^+$ & 12-11 & & & & &  35.669 & 31.8$\pm$ 5.4&  \\
36.268 & 275.73 & HNC & 12-11 & & & 36.266 & 144.0$\pm$ 6.9&  & &  \\
36.604 & 273.19 & o-H$_2$O & 3$_{12}$-3$_{03}$ & 36.586 & 157.8$\pm$ 15.6& 36.601 & 276.0$\pm$ 7.5&  & &  \\
36.737 & 272.20 & $^{13}$CO & 10-9 & 36.736 & 479.4$\pm$ 18.0& 36.737 & 166.8$\pm$ 3.0&  6.730 & 77.4$\pm$ 4.2 \\
37.137 & 269.27 & p-H$_2$O & 1$_{11}$-0$_{00}$ & 37.140 & 183.3$\pm$ 6.9& 37.138 & 257.1$\pm$ 4.2&  37.119 & 42.9$\pm$ 11.7&  \\
37.416 & 267.27 & H$^{13}$CN & 13-12 & 37.425 & 258.9$\pm$ 9.6& 37.423 & 110.4$\pm$ 6.6&  & &  \\
37.451 & 267.01 & CN & {\it N}=10-9;$\nu$=1 & & & & &  37.448 & 21.6$\pm$ 10.2&  \\
37.800 & 264.55 & CN & {\it N}=10-9;$\nu$=0 & & & 37.799 & 171.6$\pm$ 3.3&  & &  \\
37.845 & 264.24 & CCH & {\it N}=13-12 & & & & &  37.842 & 35.4$\pm$ 11.1&  \\
38.076 & 262.63 & $^{12}$CO & $\nu$=1;10-9 & 38.083 & 67.5$\pm$ 72.3& & &  & &  \\
38.247 & 261.46 & o-H$_2$O & 7$_{25}$-8$_{18}$ & 38.235 & 16.5$\pm$ 15.3& & &  & &  \\
38.408 & 260.36 & HCN & 13-12 & 38.421 & 4116.0$\pm$ 23.4& 38.424 & 2732.0$\pm$ 20.1&  38.416 & 2275.0$\pm$ 10.8&  \\
+38.426 & 260.24 & $^{12}$CO & 10-9 & \\
38.597 & 259.09 & HCN & $\nu_2$=1;13-12f & 38.597 & 140.4$\pm$ 24.9& & &  & &  \\
38.638 & 258.82 & o-H$_2$O & 6$_{34}$-5$_{41}$ & & & 38.638 & 131.1$\pm$ 19.2&  & &  \\
+38.651 & 258.73 & HCO$^+$ & 13-12 & & &\\ 
38.791 & 257.79 & o-H$_2$O & 3$_{21}$-3$_{12}$ & & & 38.793 & 223.2$\pm$ 6.3&  & &  \\
38.972 & 256.59 & o-H$_2$O & 8$_{54}$-7$_{61}$ & 38.969 & 250.2$\pm$ 6.9& 38.966 & 205.2$\pm$ 6.9&  & &  \\
39.286 & 254.55 & HNC & 13-12 & 39.293 & 75.0$\pm$ 5.7& 39.279 & 139.5$\pm$ 5.1&  & &  \\
40.282 & 248.25 & p-H$_2$O & 4$_{22}$-4$_{13}$ & 40.280 & 285.0$\pm$ 3.9& 40.278 & 113.1$\pm$ 4.2&  & &  \\
+40.290 & 248.20 & H$^{13}$CN & 14-13 & & & \\
40.406 & 247.49 & $^{13}$CO & 11-10 & 40.401 & 479.7$\pm$ 4.2& 40.402 & 185.7$\pm$ 4.5&  40.392 & 54.9$\pm$ 4.2&  \\
40.517 & 246.81 & p-H$_2$O & $\nu_2$=1;3$_{12}$-3$_{03}$ & & & & &  40.510 & 5.7$\pm$ 3.9&  \\
40.530 & 246.73 & p-H$_2$O & $\nu_2$=1;8$_{45}$-7$_{52}$ & 40.525 & 356.1$\pm$ 3.6& 40.524 & 220.2$\pm$ 3.9&  & &  \\
40.750 & 245.40 & CCH & {\it N}=14-13 & & & 40.752 & 64.5 $\pm$ 4.2 & & &  \\
40.988 & 243.97 & p-H$_2$O & 2$_{20}$-2$_{11}$ & & & 40.992 & 102.0$\pm$ 3.9&  40.981 & 31.5$\pm$ 4.5&  \\
+40.996 & 243.93 & H~{\sc i} & H17$\alpha$ & & \\
41.191 & 242.77 & CN & {\it N}=11-10;$\nu$=1 & & & 41.201 & 21.3$\pm$ 3.0&  & &  \\
41.358 & 241.79 & HCN & 14-13 & 41.356 & 876.0$\pm$ 11.1& 41.357 & 600.9$\pm$ 3.9&  41.358 & 16.5$\pm$ 3.6&  \\
41.561 & 240.61 & HCN & $\nu_2$=1;14-13f & 41.545 & 166.2$\pm$ 11.4& & &  & &  \\
42.263 & 236.61 & $^{12}$CO & 11-10 & 42.263 & 3765.0$\pm$ 24.0& 42.265 & 2615.0$\pm$ 13.2&  42.253 & 2365.0$\pm$ 12.6&  \\
43.162 & 231.68 & H$^{13}$CN & 15-14 & 43.156 & 114.6$\pm$ 9.0& & &  & &  \\
43.655 & 229.07 & CCH & {\it N}=15-14 & & & & &  43.649 & 29.7$\pm$ 3.0&  \\
44.073 & 226.90 & $^{13}$CO & 12-11 & 44.072 & 512.7$\pm$ 10.5& 44.078 & 262.8$\pm$ 3.3&  44.062 & 64.5$\pm$ 3.0&  \\
44.308 & 225.69 & HCN & 15-14 & 44.303 & 852.3$\pm$ 10.2& 44.306 & 570.3$\pm$ 5.4&  & &  \\
44.525 & 224.59 & HCN & $\nu_2$=1;15-14f & 44.523 & 133.5$\pm$ 22.8& & &  & &  \\
44.919 & 222.62 & CN & {\it N}=12-11;$\nu$=1 & 44.903 & 20.1$\pm$ 9.3& & &  & &  \\
45.318 & 220.66 & HNC & 15-14 & 45.324 & 65.4$\pm$ 7.5& 45.323 & 160.5$\pm$ 3.9&  & &  \\
46.098 & 216.93 & $^{12}$CO & 12-11 & 46.096 & 4356.0$\pm$ 15.3& 46.097 & 3009.0$\pm$ 100.5&  46.087 & 2370.0$\pm$ 9.3&  \\
46.428 & 215.39 & HN$^{13}$C & 16-15 & 46.443 & 32.7$\pm$ 16.2& 46.439 & 20.4$\pm$ 6.9&  & &  \\
46.922 & 213.12 & p-H$_2$O & $\nu_2$=1;3$_{21}$-3$_{12}$ & & & 46.918 & 7.5$\pm$ 4.8&  & &  \\
47.053 & 212.53 & o-H$_2$O & 5$_{23}$-5$_{14}$ & & & 47.049 & 43.8$\pm$ 6.6&  & &  \\
47.256 & 211.61 & HCN & 16-15 & 47.255 & 837.9$\pm$ 9.9& 47.255 & 570.0$\pm$ 7.5&  & &  \\
47.487 & 210.58 & HCN & $\nu_2$=1;16-15f & & & 47.469 & 73.8$\pm$ 12.6&  & &  \\
47.555 & 210.28 & HCO$^+$ & 16-15 & & & 47.555 & 73.2$\pm$ 11.7&  47.565 & 28.2$\pm$ 5.4&  \\
47.738 & 209.48 & $^{13}$CO & 13-12 & 47.737 & 488.1$\pm$ 16.2& 47.740 & 217.2$\pm$ 1.8&  47.726 & 80.4$\pm$ 6.3&  \\
48.333 & 206.90 & HNC & 16-15 & & & 48.329 & 91.8$\pm$ 1.5&  48.337 & 20.7$\pm$ 6.3&  \\
48.655 & 205.53 & CN & {\it N}=13-12;$\nu$=1 & & & 48.658 & 118.5$\pm$ 3.6&  & &  \\
48.738 & 205.18 & [N~{\sc ii}] & 2-1 & & & & &  48.725 & 73.5$\pm$ 7.8&  \\
48.921 & 204.41 & H~{\sc i} & H16$\alpha$ & 48.924 & 212.4$\pm$ 22.5& 48.922 & 177.6$\pm$ 3.3&  48.918 & 102.3$\pm$ 7.2&  \\
49.117 & 203.60 & CN & {\it N}=13-12;$\nu$=0 & & & 49.121 & 219.9$\pm$ 4.5&  & &  \\
49.932 & 200.27 & $^{12}$CO & 13-12 & 49.932 & 4329.0$\pm$ 27.3& 49.933 & 3387.0$\pm$ 15.3&  49.921 & 2252.0$\pm$ 9.6&  \\
50.203 & 199.19 & HCN & 17-16 & 50.204 & 709.2$\pm$ 22.5& 50.205 & 552.3$\pm$ 6.6&  & &  \\
50.521 & 197.94 & HCO$^+$ & 17-16 & & & 50.515 & 117.6$\pm$ 14.1&  & &  \\
51.402 & 194.55 & $^{13}$CO & 14-13 & 51.393 & 444.9$\pm$ 22.5& 51.381 & 298.5$\pm$8.4&  & &  \\

%% file: 14589_t2.tex
15.657 & 638.69 &  201.3 $\pm$    9.0 & NH$_2$ \\
15.762 & 634.44 &  139.5 $\pm$    8.7 & H$^{13}$CN v2 ; $^{30}$SiS \\
15.931 & 627.71 &  116.1 $\pm$    8.4 & SiO 11-10 \\
16.547 & 604.34 &  193.8 $\pm$    6.3 & NH$_2$ \\
16.675 & 599.70 &  122.7 $\pm$    5.1 & H$^{13}$CN v1+v2 \\
16.856 & 593.26 &  304.8 $\pm$   12.0 & o-SH$_2$ \\
16.998 & 588.30 &  193.8 $\pm$   21.9 & CH$_2$ \\
17.927 & 557.82 &  194.1 $\pm$    8.4 & CH \\
18.766 & 532.88 &  122.1 $\pm$    3.9 & SiS ; SiO 13-12 \\
18.945 & 527.84 &  195.0 $\pm$    3.9 & p-SH$_2$ \\
19.440 & 514.40 &  159.9 $\pm$    3.6 &  \\
20.122 & 496.97 &  130.5 $\pm$    7.5 &  \\
21.208 & 471.52 &  201.6 $\pm$   12.0 &  \\
29.287 & 341.45 &  141.9 $\pm$    4.8 & HCN v2+2v3 ; C$_{18}$O \\
31.479 & 317.67 &  134.7 $\pm$   20.4 &  \\
31.778 & 314.68 &  114.9 $\pm$   20.4 & NH$_2$ \\
32.665 & 306.14 &  150.3 $\pm$    7.8 & HCN $\nu_2$=0 \\
36.432 & 274.48 &  109.5 $\pm$   15.6 &  \\
38.842 & 257.45 &  105.9 $\pm$    4.8 &  \\
39.685 & 251.98 &  100.5 $\pm$    6.0 &  \\
47.456 & 210.72 &  101.7 $\pm$   72.3 & HCN $\nu_2$=1 \\
47.522 & 210.43 &  128.1 $\pm$   85.5 &  \\
49.589 & 201.66 &  146.1 $\pm$   27.0 &  \\
50.814 & 196.80 &  114.9 $\pm$   22.2 & HCN 3v2 \\

%% file: 14589_t3.tex
15.782 & 633.63 &  111.0 $\pm$   18.9 & H$^{13}$CN v2 \\
16.472 & 607.09 &  129.3 $\pm$    4.8 &  \\
18.461 & 541.68 &  206.7 $\pm$   12.3 &  \\
19.444 & 514.30 &  116.1 $\pm$    5.4 &  HCN maser? \\
19.579 & 510.75 &  110.7 $\pm$    5.4 &  \\
27.396 & 365.02 &  150.3 $\pm$    2.7 &  \\
29.715 & 336.53 &  189.6 $\pm$    8.1 &  HCN 2v2 \\
36.418 & 274.59 &  105.9 $\pm$   28.8 &  \\

%% file: 14589_t4.tex
15.650 & 638.98 &  201.9 $\pm$    9.0 & NH$_2$ \\
16.983 & 588.82 &  128.1 $\pm$   20.1 & CH$_2$ \\
17.662 & 566.19 &  249.3 $\pm$   30.6 & NH$_2$ \\
17.880 & 559.28 &  101.1 $\pm$   10.8 & CH \\
17.941 & 557.38 &  120.9 $\pm$   13.2 & H37$\epsilon$ \\
18.062 & 553.65 &  121.5 $\pm$    5.1 & H$^{+}$, He$^{+}$, C$^+$, S$^+$ 28$\beta$ \\
18.939 & 528.01 &  181.8 $\pm$   11.1 & H34$\delta$; p-SH$_2$ \\
20.101 & 497.49 &  109.8 $\pm$   16.2 & H$^{+}$, He$^{+}$, C$^+$, S$^+$ 27$\beta$ \\
48.138 & 207.74 &  100.5 $\pm$    7.5 & NH$_2$, H26$\epsilon$ \\